# Efficiency Gains from Using Auxiliary Variables in Imputation

Paul von Hippel (University of Texas, Austin)

Jamie Lynch (St. Norbert College)

*Correspondence*: paulvonhippel.utaustin@gmail.com

## Abstract

Imputation models sometimes use auxiliary variables that, though not part of the planned analysis, can improve the accuracy of imputed values and the efficiency of point estimates. A recent article, using evidence from simulations, argued that the use of auxiliary variables in imputation did not improve efficiency. We review the simulation results and find that the use of auxiliary variables did improve efficiency; under some conditions the efficiency gain was equivalent to increasing the sample size by a quarter. We give an example from our own research where the efficiency gained from auxiliary variables was equivalent to increasing the sample size by three quarters, and pushed some estimates from statistical insignificance to significance. For auxiliary variables to make a difference, there must be a lot of missing data, some estimates must be near the border of significance, and the auxiliary variables must be excellent predictors of the missing values.

# INTRODUCTION

Multiple imputation (MI) is a widely used method for analyzing data with missing values. Under MI, we specify an *imputation model* that fills in missing values with random imputations. After imputing missing values, we analyze the imputed data using an *analysis model*. The process of imputation and analysis are repeated $M$ times, and the estimates obtained from the $M$ analyses are combined using appropriate formulas (Rubin 1987).

The imputation model should use, at a minimum, all of the variables that will be used in the analysis model. For example, if the intended analysis is a regression of $Y$ on $X$, then the model for imputing $X$ should include $Y$, and the model for imputing $Y$ should include $X$. In addition, it is sometimes helpful for the imputation model to include one or more *auxiliary variables Z* that are not part of the analysis model but can improve the imputation of $X$ or $Y$. The use of auxiliary variables in the imputation model can increase the efficiency of estimates obtained from the analysis model (Collins, Schafer, and Kam 2001; Meng 1994).

In a recent issue of *Sociological Methods and Research*, Mustillo (2012) presented simulation results which suggested to her that "the inclusion of any type of auxiliary variable does not appreciably impact…efficiency in this simulation, regardless of the amount of missing data. Hence, the inclusion of auxiliary variables may not be necessary in many analytic situations" (page 335).

In this comment, we show that auxiliary variables did in fact improve efficiency in Mustillo's simulation. In fact, the efficiency gains were as large as could be hoped for under the simulated conditions. Under some conditions the efficiency gains were equivalent to increasing the sample size by more than a quarter. While these efficiency gains did not affect the statistical significance of Mustillo's regressions, we give an example from our research where the efficiency gained from auxiliary variables push some estimates from statistical insignificance to significance.

While we do not think that auxiliary variables should be used routinely in imputation, there are circumstances when they can be quite helpful. Those circumstances occur when there are a lot of missing values, the auxiliary variables are excellent predictors of the missing values, and there are borderline-significant results that might benefit from some increase in efficiency.

# MUSTILLO'S SIMULATION

Mustillo drew $n = 1,810$ complete cases from the General Social Survey and used them to fit a logistic regression of financial satisfaction $Y$ (1=satisfied, 0=dissatisfied) on years of education $X_1$, income $X_2$, sex $X_3$, age $X_4$ race/ethnicity $X_5$, number of children $X_6$, and marital status $X_7$. Her primary focus was on estimating the slope $\beta_1$ of education $X_1$.

To explore the implications of missing values, Mustillo deleted $X_1$ values from some of the cases. She varied the percentage of $X_1$ values that were missing ($10\%, 20\%, 30\%$), and also varied whether they were missing at random (MAR) or missing completely at random (MCAR). In this context, MCAR means that the probability of deleting an $X_1$ value does not depend on other variables in the dataset, while MAR means that the probability of deleting an $X_1$ value can depend on the fully observed $X$ variables or on $Y$ (Rubin 1976).

We will begin with the condition where 30% of $X_1$ values were MCAR. In this situation it is useful to consider the properties of the simplest estimation method, which is *listwise deletion* (LD). Under LD, we would simply delete the 30% of cases with missing $X_1$ values, and fit the logistic regression to the 70% of cases that are complete. The resulting estimates are approximately[1] unbiased since, with values missing completely at random, the complete cases are a random sample of all the cases. However, the reduction in sample size entails a loss of efficiency. Since the LD sample were 30% smaller than the complete-data sample—or, equivalently, the complete-case sample is 43% larger than the LD sample ($.43 = 1/(1 - .3) - 1$)—the complete-data standard error is 16% smaller than the standard error obtained under LD ($.16 = 1 - \sqrt{.7}$).

Let us now consider the estimates that Mustillo obtained by MI. Mustillo deleted values from the GSS 100 times to create 100 incomplete datasets. In each incomplete dataset, she imputed values $M$=50 times and obtained an MI estimate $\hat{\beta}_1$. Mustillo's imputation model was a linear regression of $X_1$ on other variables. In her most restricted imputation model, the regressors were just the variables in the analysis model—namely $X_2, X_3, ..., X_7$, and $Y$. She then specified two imputation models that used auxiliary variables; in one model the auxiliary variables were moderately predictive, and in the other they were strongly predictive.

The mean and standard deviation of the MI estimate $\hat{\beta}_1$ across the 100 simulations were used to estimate the expectation and standard error of the MI estimate. The results for each of the three imputation models are summarized in Table 1a.

Looking only at the MI estimates, which are what Mustillo presented in her tables, it is easy to get the impression that the auxiliary variables made little difference. With or without auxiliary variables, the standard error is about .02 and the expected slope is about .12. This is probably why Mustillo concluded that "including auxiliary variables did not appear to systematically…increase efficiency" (page 355).

---

[1] Logistic regression estimates are biased in small samples (King and Zeng 2001); however, in this simulation, with $n = 1,810$ and a reasonable balance of zeroes and ones, the bias is negligible.

Our Table 1a, however, puts these results in context by providing the estimate that Mustillo reported for the complete data (before deletion of values), as well as the estimate that she would have obtained under LD. These estimates provide an upper and lower bound for what we can expect from MI. Analysts typically hope that MI estimates will be more efficient than LD, but we cannot hope that MI estimates will be as efficient as estimates from complete data.

In this example, the standard errors of the complete-data and LD estimates differ by 16%, so the various MI standard errors cannot differ by more than that. With those bounds in mind, the increase in efficiency that comes from auxiliary variables looks pretty good. Without auxiliary variables, the standard error of the MI estimate is only 2% smaller than the LD standard error, but with strongly predictive auxiliary variables the standard error of the MI estimate is fully 13% smaller than the LD standard error and only 3% larger than the complete-data standard error. The efficiency gains are as much as one could hope for under the simulated conditions. The efficiency of the MI estimate with auxiliary variables is equivalent to increasing the LD sample size by 31%.

Tables 1b-c give Mustillo's results for the settings where 20% or 10% of $X_1$ values were missing at random. The results show the same pattern. Without auxiliary variables, the MI estimates are comparable to the LD estimates, but with auxiliary variables the MI estimates are nearly as efficient as the estimates from complete data. The efficiency gained from auxiliary variables is small in an absolute sense, but the gains are large given what is possible under the simulated conditions.

These results are not surprising. The efficiency to be gained from MI depends in part on the imputation model's accuracy in predicting the missing values. A perfectly accurate imputation model would have $R^2 = 1$; such a model would effectively restore the deleted $X_1$ values, so that the MI estimates would be as efficient as if no values had been missing in the first place. None of Mustillo's imputation models has $R^2 = 1$, of course, but the model with strong auxiliaries has $R^2 = .62$ and produces estimates that are almost as accurate as if no data were missing. The model without auxiliary variables, by contrast, has $R^2 = .14$ and the resulting estimates are little better than those obtained by just deleting the incomplete cases.

### An aside on listwise deletion

We should say something about the differences between the LD estimates and the MI estimates in Mustillo's simulation. With 10% or 20% of $X_1$ values missing, and no auxiliary variables in the imputation model, the MI standard errors were not smaller than the LD standard error; instead, the MI standard errors were 1% larger. This could be due to Monte Carlo error in the simulation, or it could mean that the small amount of information that MI recovered from the incomplete cases was outweighed by the random variation that MI added to the estimates through the imputed values. Auxiliary variables improved the relative efficiency of MI. When even moderately predictive auxiliary variables were added to the imputation model, the MI standard errors were smaller than the LD standard errors.

The finding that LD is comparable to MI without auxiliary variables is not true in general. The conditions of Mustillo's simulation were favorable to LD since values were missing from only one variable and since the imputation model, without auxiliary variables, was not particularly accurate at imputing the missing values. If values were missing from different variables in different cases, then LD would have to delete a larger fraction of cases and its efficiency would suffer compared to that of MI. In addition, if the imputation model were more accurate without auxiliary variables, the efficiency of the no-auxiliary MI estimates would improve.

A final point is that the $X_1$ values in Table 1 are MCAR, and under the MCAR assumption LD estimates are consistent. In another part of Mustillo's simulation, $X_1$ values were MAR but the probability that $X_1$ was missing depended only on the values of other $X$ variables; under this setting, too, LD estimates are consistent. There are situations, though, where LD is biased. In particular, LD is biased and inconsistent when deletion of $X_1$ values depends on $Y$. The bias comes from the fact that, after deletion, the sample has been selected on $Y$ (Little and Rubin 2002). In that situation, MI estimates remain consistent and hold an advantage over LD estimates with respect to bias, aside from any differences in efficiency.

## WHEN DO AUXILIARY VARIABLES MAKE A DIFFERENCE?

Mustillo's simulation makes some excellent points. First, it suggests that the efficiency to be gained from using auxiliary variable can be small unless the auxiliary variables are very good predictors of the variables to be imputed. In Table 1, the imputation model with moderately predictive auxiliary variables is only a little more efficient than the imputation model with no auxiliaries at all. Only the imputation model with strongly predictive auxiliary variables provides a substantial increase in efficiency. Past simulations have found this as well. For example, Collins et al. (2001) found that standard errors shrank very little if the auxiliary variable had a .4 correlation with the incomplete variable. However, if the auxiliary variable had a .9 correlation with the incomplete variable, then standard errors could be reduced by 20% or more (e.g., Collins et al. 2001, Table 5).

A second important point is that auxiliary variables are unlikely to help much unless there is a substantial amount of missing data. Table 1a shows the largest efficiency gains because 30% of $X_1$ values are missing. In Tables 1b-c, where just 10% or 20% of $X_1$ values are missing, relatively little efficiency was lost when values were deleted, so relatively little can be regained by using auxiliary variables. This point applies not just to auxiliary variables, but to MI in general. Schafer (1997, page 1), a leading proponent of MI, has written that "when the incomplete cases comprise only a small fraction of all cases (say, five percent or less) then [listwise] deletion may be a perfectly reasonable solution to the missing-data problem." When more cases are incomplete, however, MI can be helpful, and the advantages of MI can be considerably enhanced by adding strong auxiliary variables to the imputation model.

A final point is that there is not much need to increase efficiency if the standard errors are already small enough for the researcher's purposes. In Mustillo's simulation, the expected point estimate is 5 or 6 times its standard error, so that the null hypothesis $H_0: \beta_1 = 0$ would be rejected at any conventional significance level. A 15% increase in efficiency simply doesn't matter under those circumstances.

There are circumstances, however, where strong auxiliary variables are readily available and the associated efficiency gains can make a big difference. Longitudinal studies provide good opportunities to use auxiliary variables, because the variables observed in one round of the survey are often excellent predictors of the same variables observed in earlier and later rounds. An example occurs in the 1997 cohort of the National Longitudinal Study of Youth (NLSY97), which began with participants who were 12 to 16 years old on December 31, 1996 and then interviewed the same participants every year from 1997 until 2009.[2] In our applied research, we have estimated the regression of educational attainment (highest degree received) on BMI at age 27. Missing values pose a challenge because half of respondents had not reached age 27 by the time of the 2009 interview. We could not afford to delete half the sample, especially since we planned to conduct analyses on relatively small subsets of respondents, such as Hispanic females.

The first column of Table 2 shows LD estimates for the regression of BMI $Y$ on the educational attainment $X$ of Hispanic females at age 27. The analysis has limited power because it is limited to the 416 Hispanic females with observed values for both $X$ and $Y$. The limits to power are illustrated most clearly by the coefficient for receipt of a graduate degree. The estimated coefficient is quite large, suggesting that Hispanic females with a graduate degree are fully 4.7 kg/m$^2$ lighter than Hispanic females who dropped out of high school. For a woman of average height (164 cm), 4.7 kg/m$^2$ is almost 13 kg, or 28 lb—a substantial difference. Yet despite its size, the coefficient is not quite significant ($.05 < p < .10$), partly because of the limited number of observed values and partly because only 2% of Hispanic females received graduate degrees.

In an attempt to increase the effective sample size, we imputed missing values $M = 50$ times using the *mi impute chained* command in Stata 12. This command implements imputation by chained equations (van Buuren and Groothuis-Oudshoorn 2010), in which each variable is imputed by iteratively regressing it on all the other variables. The chained model used ordinal regression to impute educational attainment, and used linear regression to impute BMI.

---

[2] An additional interview was conducted in 2011, but the 2011 data had not been released at the time when we conducted these analyses.

Our simplest imputation model does not use auxiliary variables and simply regresses $Y$ on $X$ and $X$ on $Y$. This approach offers little help, because most of the cases that are missing $Y$ are missing $X$ as well, and because the imputation models are limited in accuracy. The imputation model for $Y$ has an $R^2$ of only .02 and the imputation model for $X$ has a pseudo-$R^2$ of only .01. The second column of Table 2 shows the resulting MI estimates. Compared to LD, imputation has increased the number of available cases by only 7%, and reduced the standard errors by only 1%. The coefficient for possession of a graduate degree remains not quite significant ($.05 < p < .10$).

In this imputation model, the obvious auxiliary variables $Z$ to use are lagged measures of $X$ and $Y$. That is, we supplemented the imputation model for educational attainment and BMI at age 27 with the same variables at ages 26, 25, and 24. The imputation model actually fit too well and predicted education status perfectly at age 27. Since perfect prediction can cause estimation problems, Stata failed to fit the imputation model until we dropped age 26 educational attainment as a regressor. After that change, the imputation model for $Y$ had an $R^2$ of .84 and the imputation model for $X$ had a pseudo-$R^2$ of .63. That is, both imputation models were much more predictive than they were without auxiliary variables.

After adding the auxiliary variables, we re-imputed the data and obtained the estimates in the last column of Table 2. Many more cases became available for analysis, and the standard errors shrank by up to 24% compared to the LD standard errors—a gain in efficiency that is equivalent to increasing the sample size by three-quarters. The efficiency gains are enough to put an asterisk on the coefficient for a graduate degree, whose significance declines from $p<.10$ to $p<.05$. The coefficient for a bachelor's degree also gains an asterisk; its significance declines from $p<.05$ to $p<.01$. These are meaningful changes to the estimates.

←Table 2 near here→

## CONCLUSION

The previous example shows that strong auxiliary variables can sometimes make a substantive difference in MI. Auxiliary variables can push borderline estimates from insignificance to significance when there are a lot of missing values and the auxiliary variables are highly predictive. This situation arises frequently in longitudinal data and in other situations where there are multiple measures of the same or similar variables.

Although the use of auxiliary variables can be quite helpful, it should not necessarily be routine. There are many situations where the available auxiliary variables are not particularly good, or when efficiency gains are not needed because estimates obtained without auxiliary variables are as precise as the investigator needs them to be. When good auxiliary variables are available, however, and a small increase in efficiency would be valuable, the possibility of adding auxiliary variables to the imputation model is a good trick to know.

# REFERENCE LIST

Van Buuren, Stef, and Karin Groothuis-Oudshoorn. 2010. "MICE: Multivariate Imputation by Chained Equations in R." *Journal of Statistical Software* 20(2).

Collins, Linda M., Joseph L. Schafer, and Chi-Ming Kam. 2001. "A Comparison of Inclusive and Restrictive Strategies in Modern Missing Data Procedures." *Psychological Methods* 6(4):330–51.

King, Gary, and Langche Zeng. 2001. "Logistic Regression in Rare Events Data." *Political Analysis* 9(2):137–63.

Little, Roderick J. A., and Donald B. Rubin. 2002. *Statistical analysis with missing data.* Hoboken, NJ: Wiley.

Meng, Xiao-Li. 1994. "Multiple-Imputation Inferences with Uncongenial Sources of Input." *Statistical Science* 9(4):538–58.

Mustillo, Sarah. 2012. "The Effects of Auxiliary Variables on Coefficient Bias and Efficiency in Multiple Imputation." *Sociological Methods & Research* 41(2):335–61.

Rubin, Donald B. 1976. "Inference and missing data." *Biometrika* 63(3):581 –592.

Rubin, Donald B. 1987. *Multiple imputation for nonresponse in surveys.* New York: Wiley.

Schafer, Joseph L. 1997. *Analysis of incomplete multivariate data.* London; New York: Chapman & Hall.

# TABLES

Table 1. Mustillo's MI results, alongside results for listwise deletion (LD) and complete data

a. Missing 30% of $X_1$ values, completely at random

| | | Multiple imputation | | | |
|---|---|---|---|---|---|
| | LD | No auxiliary variables | Moderate auxiliaries | Strong auxiliaries | Complete data |
| $R^2$ of imputation model | | .14 | .45 | .62 | |
| Slope of education | .1256 | .1233 | .1231 | .1238 | .1256 |
| Standard error | (.0239) | (.0234) | (.0229) | (.0209) | (.0200) |
| Difference in SE vs. LD | | -2% | -4% | -13% | -16% |
| Equivalent change in sample size vs. LD | | 4% | 9% | 31% | 43% |

b. Missing 20%

| | | Multiple imputation | | | |
|---|---|---|---|---|---|
| | LD | No auxiliary variables | Moderate auxiliaries | Strong auxiliaries | Complete data |
| $R^2$ of imputation model | | .14 | .45 | .62 | |
| Slope of education | .1256 | .1237 | .1237 | .1242 | .1256 |
| Standard error | (.0224) | (.0226) | (.0217) | (.0207) | (.0200) |
| Difference in SE vs. LD | | 1% | -3% | -7% | -11% |
| Equivalent change in sample size vs. LD | | -2% | 6% | 17% | 25% |

c. Missing 10%

| | | Multiple imputation | | | |
|---|---|---|---|---|---|
| | LD | No auxiliary variables | Moderate auxiliaries | Strong auxiliaries | Complete data |
| $R^2$ of imputation model | | .14 | .45 | .62 | |
| Slope of education | .1256 | .1243 | .1250 | .1251 | .1256 |
| Standard error | (.0211) | (.0212) | (.0207) | (.0203) | (.0200) |
| Difference in SE vs. listwise deletion | | 1% | -2% | -4% | -5% |
| Equivalent change in sample size vs. LD | | -1% | 4% | 8% | 11% |

Table 2. Regression of BMI on educational attainment among Hispanic females in the NLSY97

| | LD | MI without auxiliaries | MI with auxiliaries |
|---|---|---|---|
| N | 416 | 445 | 924 |
| $R^2$ of imputation model for Y | | .02 | .84 |
| Pseudo-$R^2$ of imputation model for X | | .01 | .63 |
| High school diploma / GED | -0.68 | -0.76 | -1.09 |
| (SE) | (1.02) | (1.01) | (0.77) |
| % reduction in SE vs. LD | | -1% | -24% |
| equivalent increase in sample size | | 1% | 74% |
| Associate's degree | -0.28 | -0.35 | -1.63 |
| (SE) | (1.60) | (1.60) | (1.36) |
| % reduction in SE vs. LD | | 0% | -15% |
| equivalent increase in sample size | | 0% | 38% |
| Bachelors' degree | -2.72* | -2.78* | -3.35** |
| (SE) | (1.29) | (1.28) | (0.98) |
| % reduction in SE vs. LD | | -1% | -24% |
| equivalent increase in sample size | | 1% | 71% |
| Graduate degree | -4.68† | -4.48† | -5.01* |
| (SE) | (2.66) | (2.63) | (2.06) |
| % reduction in SE vs. LD | | -1% | -22% |
| equivalent increase in sample size | | 2% | 66% |
| Intercept | 29.41*** | 29.49*** | 29.55*** |
| (SE) | (.92) | (.91) | (.72) |
| % reduction in SE vs. LD | | -1% | -21% |
| equivalent increase in sample size | | 2% | 61% |

†$p<.10$, *$p<.05$, **$p<.01$, ***$p<.001$